\documentclass[useAMS,usenatbib]{mn2e}

\usepackage{graphicx}

\newcommand{\apj}{ApJ}
\newcommand{\mnras}{MNRAS}

\title[]{Complex molecules in  W51 North region}

\author[Rong et al.]{Jialei Rong$^{1}$\thanks {E-mail:JiaLeiRong@163.com,slqin@bao.ac.cn}, Sheng-Li Qin$^{1\star}$, Luis A.\ Zapata$^2$, Yuefang Wu$^{3}$,
\newauthor Tie Liu$^{4}$, Chengpeng Zhang$^{3}$, Yaping Peng$^{1}$, Li Zhang$^{1}$ and Ying Liu$^{5}$ \\
 $^{1}$Department of Astronomy, Yunnan University, and Key Laboratory of Astroparticle Physics of Yunnan Province,\\
  Kunming, 650091, China \\
$^{2}$Centro de Radiostronom\'{\i}a y Astrof\'{\i}sica, Universidad Nacional Aut\'onoma de M\'exico, 58089 Morelia, Michoac\'an, M\'exico \\
$^{3}$Department of Astronomy, Peking University, Beijing, 100871, China\\
$^{4}$Korea Astronomy and Space Science Institute 776, Daedeokdaero, Yuseong-gu, Daejeon, Republic of Korea 305-348\\
$^{5}$Department of Physics and Hebei Advanced film Laboratory, Hebei Normal University, Shijiazhuang 050024, China
}

\begin{document}
\date{}
\pagerange{\pageref{firstpage}--\pageref{lastpage}} \pubyear{2015}
\maketitle
\label{firstpage}
\begin{abstract}
We present Submillimeter Array (SMA) molecular line observations in two 2 GHz-wide bands
centered at 217.5 and 227.5 GHz, toward the massive star forming region W51 North.
We identified 84 molecular line transitions from 17 species and their isotopologues.
The molecular gas distribution of these lines mainly peaks in the continuum
position of W51 North, and has a small tail extending to the west, probably
associated with W51 d2.
In addition to the commonly detected nitrogen and oxygen-bearing species,
we detected a large amount of transitions of the  Acetone (CH$_3$COCH$_3$)
and Methyl Formate (CH$_3$OCHO), which may suggest that these molecules
are present in an early evolutionary stage of the massive stars.
We also found that W51 North is an ethanol-rich source.
There is no obvious difference in the molecular
gas distributions between the oxygen-bearing and nitrogen-bearing molecules.
Under the assumption of Local Thermodynamic Equilibrium (LTE),
with the XCLASS tool, the molecular column
densities, and rotation temperatures are estimated.
 We found that the oxygen-bearing molecules have considerable higher column densities
and fractional abundances than the nitrogen-bearing molecules.
The rotation temperatures range from 100 to 200 K,
suggesting that the molecular emission could be originated from a warm environment.
 Finally, based on the gas distributions, fractional abundances and the rotation temperatures,
we conclude that  CH$_3$OH, C$_2$H$_5$OH, CH$_3$COCH$_3$ and
CH$_3$CH$_2$CN might be synthesized on the grain surface, while gas phase
chemistry is responsible for the production of  CH$_3$OCH$_3$,
CH$_3$OCHO and  CH$_2$CHCN.
\end{abstract}



\section{Introduction}
\label{sec:intro}

There are three promising theoretical models to explain the formation of the massive stars:
monolithic collapse, competitive accretion, and mergers of low mass stars (see for a review,
Zinnecker \& Yorke 2007). However,  their relative large distances (a few parsecs), the clustered
formation environments, and their short timescales have made extremely difficult to discard any of these
models (Zapata et al. 2015).

Massive stars have a substantial impact on their surrounding environments,
making important contributions to the chemistry enrichment of the interstellar medium
(e.~g., Hern\'{a}ndez-Hern\'{a}ndez et al. 2014). Thus, searching for
complex molecules in massive star formation regions is
a crucial building block to understand massive star formation, since these molecules can
provide information on the physical conditions and evolutionary phases of
massive star formation (Herbst \& van Dishoeck 2009).

The W51 North is one of well-studied massive star-forming regions within
the luminous cluster W51-IRS2 (Zapata, Tang \& Leurini 2010). The distance
from W51 North to the Sun is approximately 7 kpc (Imai et al. 2002). However,
a more recent and accurate estimation from Xu et al. (2009) obtained
5.1$^{+2.9}_{-1.4}$ kpc.

A large number of H$_2$O and OH masers, bright dust emission, outflows, and
 infalling gas were observed toward W51 North region, indicating that this
region is forming indeed massive stars (Downes et al. 1979; Zhang, Ho \& Ohashi
1998; Zapata et al. 2008, 2009; Zapata, Tang \& Leurini 2010). Absence of
 of centimeter emission suggests that W51 North may represent an
extremely early stage of the massive star formation (Downes et al. 1979;
Zhang, Ho \& Ohashi 1998).
As the Submillimeter Array (SMA\footnote {The Submillimeter Array is a joint project
between the Smithsonian Astrophysical Observatory and the Academia
Sinica Institute of Astronomy and Astrophysics and is funded by
the Smithsonian Institution and the Academia Sinica.}) has a broad
bandwidth (4 GHz, at the time of these observations) one can detect multiple complex
molecules simultaneously,  allowing to trace the physical and chemical properties of massive star formation
environments close to the massive protostars.

In this paper, we present the results from the SMA observations of W51
North region. We have identified 17 species and obtained their physical parameters
by the use of XCLASS program. We describe the observations in \S 2. In \S 3 we
present observational results, followed by data analysis in \S4. \S 5
discusses individual molecules. We summarize the results in \S 6.


\section{OBSERVATIONS}

Track-sharing observations toward W51 North and W51 Main sources were carried
out with the SMA in 2005 August, using seven antennas in its compact array. The
phase-tracking center of W51 north was placed at
R.A.= $19^{\rm h}23^{\rm m}40^{\rm s}.05$,
decl.=$ + 14^{\circ}31^{\prime}05^{\prime\prime}.59$ (J2000.0). The typical system
temperature of 183 K indicates good weather during observations. The
observations covered frequencies from 216.5 to 218.5 GHz (lower sideband), and
226.5 to 228.5 GHz (upper sideband) with a frequency resolution of 0.8125
MHz.

The calibration and imaging were done in Miriad. Bandpass ripples were
corrected with the QSO 3C454.3 and Uranus. We also corrected baseline-based
bandpass errors using the QSO 3C454.3. QSOs 1741-038  and 1749+096 were used for
phase calibration. Uranus was used for flux calibration, and the absolute flux
scale is estimated to be accurate to within 20\%. Continuum subtraction was
made in UV domain. Self-calibration was performed to the continuum data, and
the gain solutions from the continuum were applied to the line data.

The synthesized beam sizes of the continuum and line images are approximately
3$^{\prime\prime}$.1$\times$ 2$^{\prime\prime}$.8 (P.A.$=
79.0^{\circ}$). 1 $\sigma$ rms noises of the continuum and lines are
approximately 0.01 Jy beam$^{-1}$ and 0.1 Jy beam$^{-1}$, respectively.
1 Jy beam$^{-1}$ corresponds to a main beam brightness temperature of 2.7 K.

\section{RESULTS}
\subsection{Continuum}

Continuum image was constructed from line-free channels of LSB and
USB UV data, as shown in Figure 1. The continuum image shows a compact and strong source at the position of W51 North,
with an extension to the north, which embraces IRS2d source.
KJD3 and W51d2 are located east and west of W51 North, respectively (Zapata et al. 2008).

Two dimension Gaussian fits were made to the continuum obtaining the peak intensity and total
integrated flux of  $\sim 3.0 \pm 0.3$ Jy~beam$^{-1}$  and $\sim 6.7 \pm 0.8$ Jy, respectively.

\begin{figure*}
\includegraphics[angle=-90,width=0.8\textwidth]{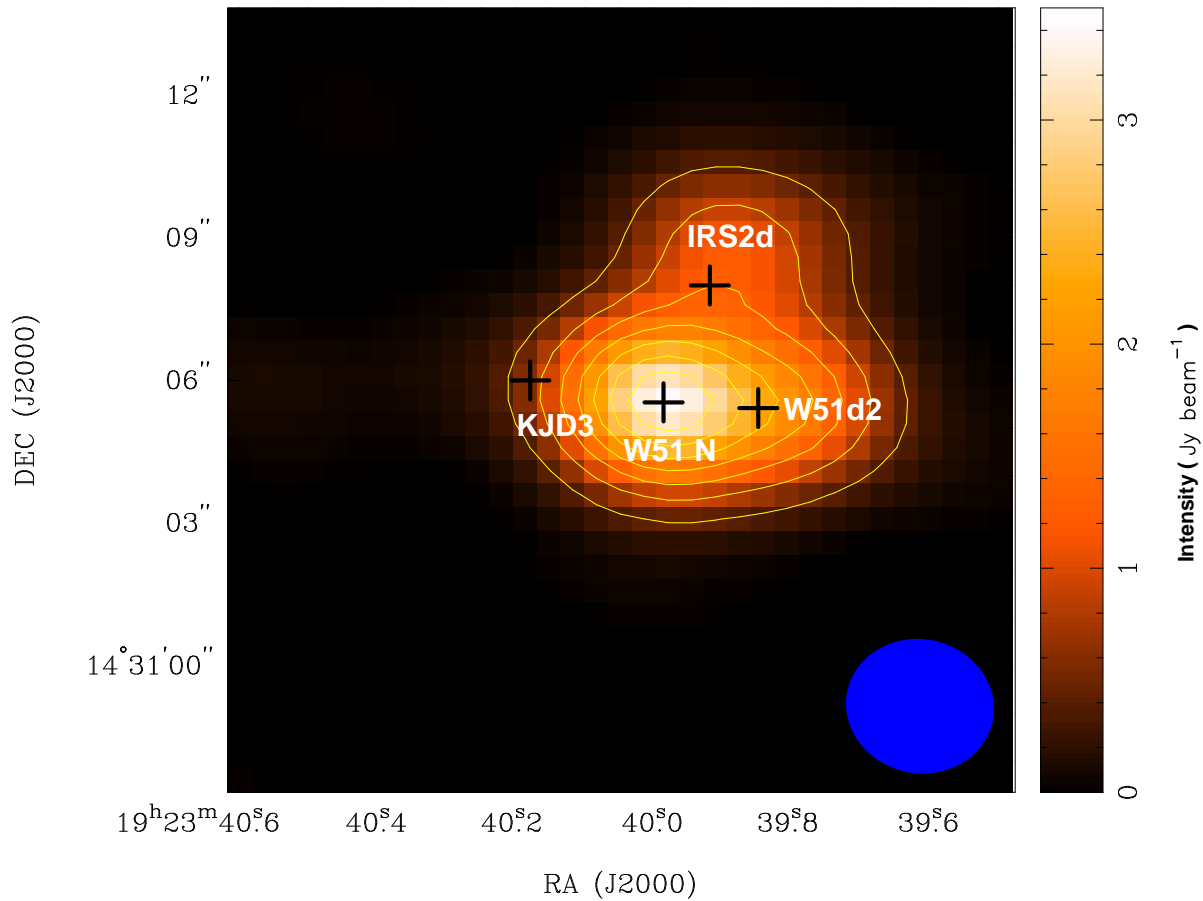}
\vspace{-0.6cm}
\caption{The continuum image at 227 GHz, shown in both contours and grey.
The contour levels are 5\%, 10\%, 20\%, 30\%, 40\%, 50\%, 60\%, 70\%,80\%,
90\%, 100\% of the maximum intensity of $\sim$3.0 Jy~beam$^{-1}$ . The color
scale on the right shows the intensity ranging from 0 to 3.5
Jy~beam$^{-1}$. The synthesized beam is shown in the bottom right corner.  }
\label{fig:F1}
\end{figure*}

Under the assumptions that an average grain radius is 0.1 $\mu$m, grain
density is 3 gr cm$^{-3}$ and a gas to dust radio is 100,
the H$_2$ column density and mass can be calculated by the
formula (Lis, Carlstrom \& Keene 1991),

\begin{equation}
N_{H_2}=8.1\times10^{17}\frac{e^{\frac{h\nu}{\kappa T}}-1}{Q(\nu)\Omega} \big(\frac{S_\nu}{Jy}\big)\big(\frac{\nu}{GHz}\big)^{-3} (cm^{-2}),
\end{equation}

and

\begin{equation}
M_{H_2}=1.3\times10^{4}\frac{e^{\frac{h\nu}{\kappa T}}-1}{Q(\nu)} \big(\frac{S_\nu}{Jy}\big)\big(\frac{\nu}{GHz}\big)^{-3}\big(\frac{D}{kpc}\big)^{2} (M_{\odot}),
\end{equation}

where $\kappa$ and $h$ are Boltzmann constant and Planck constant,
respectively. T is the dust temperature, Q($\nu$) is the grain emissivity at
frequency $\nu$, $S_\nu$ is the total integrated flux of the continuum and
$\Omega$ is the solid angle subtended by the source, D is the distance from
the source to Sun.

The CH$_3$OH is a grain surface molecule. Here, this molecule peaks
at the position of the continuum source associated with W51 North
(see Figure 3 and discussion is section 5).
Probably the dust and gas are well coupled through collisions in W51 North region.
So we assumed that the dust temperature equals the rotation temperature of 138 K derived from the
CH$_3$OH lines. We adopt $Q(\nu)=2.2\times10^{-5}$ at 1.3 mm ($\beta=1.5$)
 (Lis \& Goldsmith 1990; Lis, Carlstrom \& Keene 1991). Therefore, we obtained
 a source-averaged H$_2$ column density of $6.4 \times 10^{24}$ cm$^{-2}$ and
 an H$_2$ mass of $\sim1348M_{\odot}$.
As the continuum peaks at W51 North, then the mass estimated from peak intensity
should correspond to gas mass of W51 North source.
The  H$_2$ mass of $\sim$ 610 $\pm$ 61 $M_{\odot}$ is
 estimated from the continuum peak intensity.

 A similar result, $\sim$ 400$M_{\odot}$ was reported  by Zhang, Ho \& Ohashi (1998), in which
 they calculated the mass of W51 North based on a dust temperature of 100 K,
 continuum peak intensity of $\sim$ 2.0 Jy~beam$^{-1}$  at
 $\nu\sim140$ GHz, D of 7 $\sim$ 8 kpc and  $\beta=1.0$. If we take $\beta=1.0$, then
 Q($\nu$) $\approx2.51\times10^{-5}$ at 1.3 mm and then the H$_2$ mass of $\sim$
 535 $\pm$ 54$M_{\odot}$ is obtained. A dust temperature larger than 100 K
 suggests that dust grains are not coagulated with ice mantles, and thus $\beta$
 $\sim$ 1.5 is a reasonable guess in W51 North (Ossenkopf \& Henning 1994).
 Therefore the difference of mass between ours and $\sim$ 400$M_{\odot}$ by Zhang, Ho \& Ohashi (1998)
 may be caused by taking different value of distance, Q($\nu$) and uncertain of flux calibration.

\subsection{Molecular lines}

The Lower Sideband and Upper Sideband spectra extracted from
the continuum peak position and related with W51 North, are shown as the black curves in Figure 2.
A few CN transitions in the USB show absorption features indicating gas infalling onto
continuum core (Zapata et al. 2008). The SiO line in emission is tracing the outflow
reported by Zapata et al. (2009).

We identify all the molecular transitions using the XCLASS
program\footnote{http://www.astro.uni-koeln.de/projects/schilke/XCLASS}.
 84 molecular transitions from 17 species and their isotopologues are
 identified  (Figure 2). In this image the
corresponding molecular names are labelled, including oxygen-bearing molecules
of CH$_3$OH, H$_2$CO, C$_2$H$_5$OH, CH$_3$OCHO, CH$_3$OCH$_3$, CH$_3$COCH$_3$,
nitrogen-bearing molecules of CN, CH$_3$CH$_2$CN, HC$_3$N,
CH$_2$CHCN, sulfur-bearing molecules of SO$_2$, H$_2$S, SiS, and
deuterated molecules of DCN and CH$_3$CCD. The quantum numbers,
frequency, $S_{ij} \mu^2$ and $E_u$ of each transition are
summarized in Table 1. Column (1) lists the species and column (2)
lists quantum number, column (3), (4) and (5)
give frequency, S$_{ij}$$\mu^2$ and E$_u$. The source size
$\theta_a$$\times$$\theta_b$, $\Delta$V and $I$ are given in
column (6), (7) and (8), respectively.

\subsection{Gas distribution}

 Figure 3 presents a sample of images from the oxygen-, nitrogen-,
and sulfur-bearing molecules, which can provide spatial distribution of specific
molecules. In general, all molecular images
show a compact structure. The peak emission of the
oxygen-bearing molecules of CH$_3$OH, CH$_3$COCH$_3$ and
CH$_3$OCHO, and nitrogen-bearing molecules of CH$_2$CHCN,
CH$_3$CH$_2$CN and DCN are coincident with the continuum peak
position. The oxygen- and nitrogen-bearing molecules are probably well mixed
in space,  which imply that they may have same chemical origin (Remijan
et al. 2004). While the emission peak of the sulfur-bearing molecules
of H$_2$S, and SO$_2$ are located east of the continuum peak. Probably the
sulfur-bearing molecules have different chemical routes and form in different
environments when compared with the oxygen- and nitrogen-bearing molecules.
In addition to the compact source structure, there is an obvious gas extension
to the west of the continuum traced by the DCN, H$_2$S, and
SiS is observed.

\begin{table*}
\centering
 \caption{The molecular line parameters}
\begin{tabular}{rllrrrrrl} \hline
Molecule  & Quantum Numbers       & Frequency          &  S$_{ij}$$\mu^2$      &   E$_u$   &  $\theta_a$$\times$$\theta_b$& $\Delta$V        & I \\
          &                       &  GHz                  &debye$^2$           &K          &arcsec$^{2}$ &kms$^{-1}$& Jy/beam\\
  \hline

 CH$_3$OH&$5(1,4)-4(2,2)$&216.94556&1.12357&55.87102&$3.2\times3.0$&$8.2\pm0.1$&$5.0\pm0.1 $\\
         &$6(1,5)-7(2,6)$,  $v_t=1$&217.29920 &4.66445&373.92517&$3.2\times3.0$&$8.6\pm0.1$&$3.9\pm0.1$\\
         &$15(6,10)-16(5,11)$, $ v_t=1$ &217.64286 &4.88652&745.60607&$3.2\times3.0$&$8.0\pm0.2$&$3.0\pm0.1$\\
         &$20(1,19)-20(0,20)$&217.88639&11.50535&508.37582&$3.2\times3.0$&$8.3\pm0.2$&$2.8\pm0.1$\\
         &$4(2,2)-3(1,2)$&218.44005&3.4766&45.45988&$3.2\times2.9$&$8.5\pm0.1$&$9.9\pm0.1$\\
         &$21(1,20)-21(0,21)$&227.09460&11.5939&557.07117&$3.1\times2.8$&$7.5\pm0.2$&$2.1\pm0.1$\\
         &$16(1,16)-15(2,13)$&227.81465&5.23842&327.23797&$3.0\times2.8$&$7.7\pm0.1$&$4.3\pm0.1$\\
         &&&&&&&\\
$^{13}$CH$_3$OH&$14(1,13)-13(2,12)$&217.04462&5.78629&254.25167&$3.2\times 3.0$&$5.2\pm0.4$&$0.9\pm0.1$\\
               &$10(2,8)-9(3,7)$&217.39955&2.68164&162.41055&$3.2\times3.0$&$4.0\pm0.9$&$0.3\pm0.1$\\
               &&&&&&&\\
H$_2$CO&$3(0,3)-2(0,2)$&218.22219&16.30796&20.9564&$3.2\times2.9$&$8.4\pm0.1$&$11.1\pm0.1$\\
        &$3(2,2)-2(2,1)$&218.47563&9.06194&68.0937&$3.2\times 2.9$&$8.0\pm0.1$&$7.1\pm0.1$\\
        &&&&&&&\\
C$_2$H$_5$OH&$8(4,5)-7(3,5)$&216.65968&5.00042&106.29995&$3.2\times3.0$&$8.5\pm1.6$&$0.3\pm0.1$\\
            &$5(1,4)-4(0,4)$&217.54815&3.63648&75.60201&$3.2\times3.0$&$5.9\pm0.6$&$0.5\pm0.1$\\
            &$5(3,3)-4(2,2)$&217.80369&5.98544&23.89298&$3.2\times3.0$&$7.3\pm0.4$&$1.1\pm0.1$\\
            &$5( 3, 2)- 4( 2, 3)$&218.46123&5.59497&23.89375&$3.2\times2.9$&$10.5\pm0.1$&$4.8\pm0.1$\\
            &$26(2,24)-26(1,25)$&226.58134&28.24034&304.69227&$3.1\times2.8$&$2.0\pm1.0$&$0.2\pm0.1$\\
            &$10(2,9)-9(1,8)$&226.66170&7.05411&51.02783&$3.1\times2.8$&$6.4\pm0.4$&$1.0\pm0.1$\\
            &$13(3,10)-12(3,9)$&227.29475&19.66545&148.56961&$3.1\times2.8$&$6.7\pm0.5$&$0.9\pm0.1$\\
            &$18(5,13)-18(4,14)$&227.60608&18.61686&175.30421&$3.1\times2.8$&$11.2\pm0.2$&$3.6\pm0.1$\\
            &$3(2,2)-2(1,2)$&227.76082&2.02053&71.44923&$3.0\times2.8$&$3.6\pm0.5$&$0.6\pm0.1$\\
            &$13(1,12)-12(1,11)$&227.89191&20.6452&140.01443&$3.1\times2.8$&$9.0\pm0.6$&$0.8\pm0.1$\\
            &$13(3,10)-12(3,9)$&228.02905&19.66581&143.89662&$3.1\times2.8$&$6.1\pm0.3$&$1.2\pm0.1$\\
            &$11(1,10)-10(2,8)$&228.28855&2.47972&118.95090&$3.0\times2.8$&$3.9\pm1.7$&$0.2\pm0.1$\\
            &$7(3,4)-7(2,6)$&228.52283&4.05832&95.90890&$3.0\times2.8$&$4.5\pm0.4$&$0.4\pm0.1$\\
            &&&&&&&\\
CH$_3$OCH$_3$&$22(4,19)-22(3,20)$ AA&217.19317&162.14491&253.41131&$3.2\times 3.0$&$6.0\pm0.3$&$1.2\pm0.1$\\
             &$23(3,21)-23(2,22)$ AA &218.49503&78.0112&263.83508&$3.2\times2.9$&$6.0\pm0.8$&$0.5\pm0.1$\\
             &$26(5,21)-26(4,22)$ EA &227.64812&84.89383&355.76623&$3.1\times2.8$&$6.0\pm1.9$&$0.2\pm0.1$\\
             &$24(3,22)-24(2,23)$ AE
&227.65439&78.79933&285.56311&$3.1\times2.8$&$4.0\pm0.5$&$0.7\pm0.1$\\

             &&&&&&&\\
CH$_3$COCH$_3$&$19(4,16)-18(3,15)$ EE&217.02251&2042.48098&115.49565&$3.2\times3.0$&$6.6\pm0.2$&$2.3\pm0.1$\\
              &$19(4,16)-18(3,15)$ AA&217.07050&765.72157&115.43113&$3.2\times3.0$&$7.1\pm0.3$&$1.5\pm0.1$\\
              &$20(2,18)- 19(3,17)$ EA&218.09145&146.20943&119.17572&$3.2\times2.9$&$7.5\pm0.4$&$1.2\pm0.1$\\
              &$20(3,18)-19(2,17)$ EE&218.12721&1200.40770&119.10113&$3.2\times2.9$&$8.3\pm0.2$&$2.4\pm0.1$\\
              &$20(3,18)-19(2,17)$ AA&218.16293&1461.76777&119.02728&$3.2\times2.9$&$7.1\pm0.3$&$1.3\pm0.1$\\
              &$12(9,3)-11(8,4) $ EE&226.81261&547.66326&66.18982&$3.2\times2.9$&$7.1\pm0.3$&$1.3\pm0.1$\\
              &$20(3,17)-19(4,16)$ AA&226.87939&817.15094&126.31957&$3.1\times2.8$&$11.2\pm2.0$&$0.3\pm0.1$\\
              &$20(3,17)-19(4,16)$
EE&226.83206&2179.59244&126.38182&$3.1\times2.8$&$7.6\pm0.3$&$1.5\pm0.1$\\
&$35(10,26)-35(9,27)$ EE&227.89467&215.20891&418.29082&$3.0\times2.8$&$7.3\pm0.3$&$1.7\pm0.1$\\
              &$21(2,19)-20(3,18)$ EA&227.90395&619.07570&130.11255&$3.0\times2.8$&$7.5\pm0.3$&$1.5\pm0.1$\\
              &$21(3,19)-20(3,18)$ EE&227.93937&672.20702&130.04037&$3.0\times2.8$&$7.2\pm0.2$&$1.9\pm0.1$\\
              &$21(2,19)-20(3,18)$ AA&227.97476&1547.25807&129.96837&$3.1\times2.8$&$9.2\pm0.2$&$2.2\pm0.1$\\
              &$12(10,3)-11(9,2)$ AE&228.50287&378.05109&68.56948&$3.0\times2.8$&$3.3\pm0.6$&$0.5\pm0.1$\\

           &&&&&&&\\
CH$_3$OCHO&$18(2,16)-17(2,15)$ E&216.83020&46.13641&105.67781&$3.2\times3.0$&$7.0\pm0.1$&$2.9\pm0.1$\\
           &$18(2,16)-17(2,15)$ A&216.83889&46.14168&105.6673&$3.2\times3.0$&$8.6\pm0.2$&$2.1\pm0.1$\\
           &$20(0,20)-19(0,19)$ E&216.96625&52.81048&111.49826&$3.2\times3.0$&$9.1\pm0.1$&$6.0\pm0.1$\\
           &$17( 4,13)-16( 4,12)$ A&217.31263&42.78270&289.95716&$3.2\times3.0$&$6.0\pm0.6$&$0.7\pm0.1$\\
           &$17( 4,13)-16( 4,12)$ E&218.10844&42.94997&289.67911&$3.2\times3.0$&$7.7\pm1.2$&$0.4\pm0.1$\\
           &$17(3,14)-16(3,13)$ E&218.28090&43.5903&99.72935&$3.2\times2.9$&$7.4\pm0.2$&$3.0\pm0.1$\\
           &$17(3,14)-16(3,13)$ A&218.29789&43.60233&99.72110&$3.2\times2.9$&$11.0\pm0.3$&$2.1\pm0.1$\\
           &$33(6,28)-33(5,29)$ A&226.58140&6.87080&357.68199&$3.1\times2.8$&$4.1\pm0.9$&$0.4\pm0.1$\\
           &$20(2,19)-19(2,18)$ E&226.71306&52.04002&120.22039&$3.1\times2.8$&$9.0\pm0.2$&$2.1\pm0.1$\\
           &$20(2,19)-19(2,18)$ A&226.71875&51.409&120.207&$3.1\times2.8$&$8.5\pm0.2$&$2.0\pm0.1$\\
           &$20(1,19)-19(1,18)$
E&226.77315&51.417&120.21536&$3.1\times2.8$&$8.9\pm0.2$&$2.0\pm0.1$\\
           &$20(1,19)-19(1,18)$ A&226.77879&52.0425&120.20297&$3.1\times2.8$&$9.5\pm0.3$&$2.0\pm0.1$\\
           &$19(2,17)-18(2,16)$ E&227.01955&48.76106&116.573&$3.1\times2.8$&$7.5\pm0.2$&$2.1\pm0.1$\\
           &$19(2,17)-18(2,16)$ A&227.02810&48.4690&116.56162&$3.1\times2.8$&$8.6\pm0.2$&$2.2\pm0.1$\\
           &$21(1,21)-20(0,20)$ E&227.56269&8.88264&122.41946&$3.1\times2.8$&$8.5\pm0.1$&$6.0\pm0.1$\\
           &$18(3,15)-17(3,14)$ E&228.21129&46.25056&297.16473&$3.0\times2.8$&$3.4\pm1.2$&$0.9\pm0.1$\\
           &&&&&&&\\

\hline
\end{tabular}
\end{table*}
\setcounter{table}{0}
\begin{table*}
\caption{Continued.}
\begin{tabular}{rllrrrrrl} \hline
Molecule  & Quantum Numbers       & Frequency          &  S$_{ij}$$\mu^2$      &   E$_u$   &  $\theta_a$$\times$$\theta_b$& $\Delta$V        & I \\
          &                       &  GHz                  &debye$^2$           &K          &arcsec$^{2}$ &kms$^{-1}$& Jy/beam\\
  \hline

DCN &$J=3-2, F=3-2$&217.23855&23.85251&20.85164&$3.2\times3.0$&$7.8\pm0.1$&$6.1\pm0.1$\\
&&&&&&&\\
CH$_3$CH$_2$CN&$24(3,21)-23(3,20)$&218.38997&350.22816&139.91753&$3.2\times2.9$&$7.0\pm0.3$&$1.3\pm0.1$\\
              &$25(3,22)-24(3,21)$&227.78097&365.32437&150.84922&$3.0\times2.8$&$6.0\pm0.3$&$1.4\pm0.1$\\
              &$25(2,23)-24(2,22)$&228.48314&368.05531&146.59165&$3.0\times2.8$&$7.0\pm0.3$&$1.3\pm0.1$\\
              &&&&&&&\\
HC$_3$N&$J=24-23$&218.32479&332.86626&130.98209&$3.2\times2.9$&$8.7\pm0.1$&$5.7\pm0.1$\\
       &$J=25-24$&227.41891&346.74084&141.89637&$3.1\times2.8$&$8.5\pm0.1$&$4.9\pm0.1$\\
       &&&&&&&\\
HCC$^{13}$CN&$J=24-23$&217.41957&334.2359&130.43896&$3.2\times3.0$&$3.8\pm0.7$&$0.4\pm0.1$\\

&&&&&&&\\
CH$_2$CHCN&$23(2,22)-22(2,21)$&216.93672&996.36042&133.93474&$3.2\times3.0$&$9.0\pm0.7$&$0.8\pm0.1$\\
          &$23(6,17)-22(6,16)$&218.40245&935.94512&203.51934&$3.2\times2.9$&$5.5\pm0.5$&$0.8\pm0.1$\\
          &$23(8,16)-22(8,15)$&218.42180&882.77691&263.80899&$3.2\times2.9$&$6.2\pm1.0$&$0.4\pm0.1$\\
          &$23(10,13)-22(10,12)$&218.52000&814.48136&341.05429&$3.2\times2.9$&$5.3\pm1.2$&$0.3\pm0.1$\\
          &$24(8,16)-23(8,15)$&227.91864&931.50638&274.74733&$3.0\times2.8$&$5.7\pm1.2$&$0.3\pm0.1$\\
          &$24(5,19)-23(5,18)$&227.96759&1002.49119&190.73697&$3.1\times2.8$&$8.3\pm0.9$&$0.5\pm0.1$\\
          &$24(3,22)-23(3,21)$&228.09054&1031.47311&156.23271&$3.0\times2.8$&$9.7\pm0.8$&$0.6\pm0.1$\\
          &$24(4,21)-23(4,20)$&228.10462&1018.75491&171.33742&$3.0\times2.8$&$9.7\pm0.8$&$0.6\pm0.1$\\
          &$24(4,20)-23(4,19)$&228.16030&1018.75515&171.34687&$3.0\times2.8$&$3.6\pm1.3$&$0.2\pm0.1$\\
          &&&&&&&\\
CH$_3$CCD&$14(3)-13(3)$&218.00760&47.28557&143.50905&$3.2\times3.0$&$3.8\pm0.6$&$0.4\pm0.1$\\
         &$14(2)-13(2)$&218.02754 &24.27496&107.38842&$3.2\times2.9$&$3.8\pm0.4$&$0.7\pm0.1$\\
         &$14(0)-13(0)$&218.04350 &24.78309&78.48706&$3.2\times2.9$&$6.6\pm0.5$&$0.7\pm0.2$\\
         &&&&&&&\\
SO$_2$&$22(2,20)-22(1,21)$&216.64330&35.25156&248.44117&$3.2\times2.9$&$11.4\pm0.1$&$5.9\pm0.1$ \\
 &&&&&&&\\
 $^{33}$SO&$6(5)-5(4)$, &217.82718&10.19653&34.67135&$3.2\times3.0$&$9.5\pm0.4$&$1.2\pm0.1$\\

 &&&&&&&\\
 H$_2$S&$2(2,0)-2(1,1)$&216.71043&2.03714&83.97932&$3.2\times3.0$&$9.9\pm0.1$&$4.3\pm0.1$\\
 &&&&&&&\\
 SiS&$12-11$&217.81766&36.12197&67.95393&$3.2\times3.0$&$5.1\pm0.4$&$0.9\pm0.1$\\

 &&&&&&&\\
\hline
\end{tabular}
\end{table*}
\begin{figure*}
\includegraphics[angle=-90,width=1\textwidth]{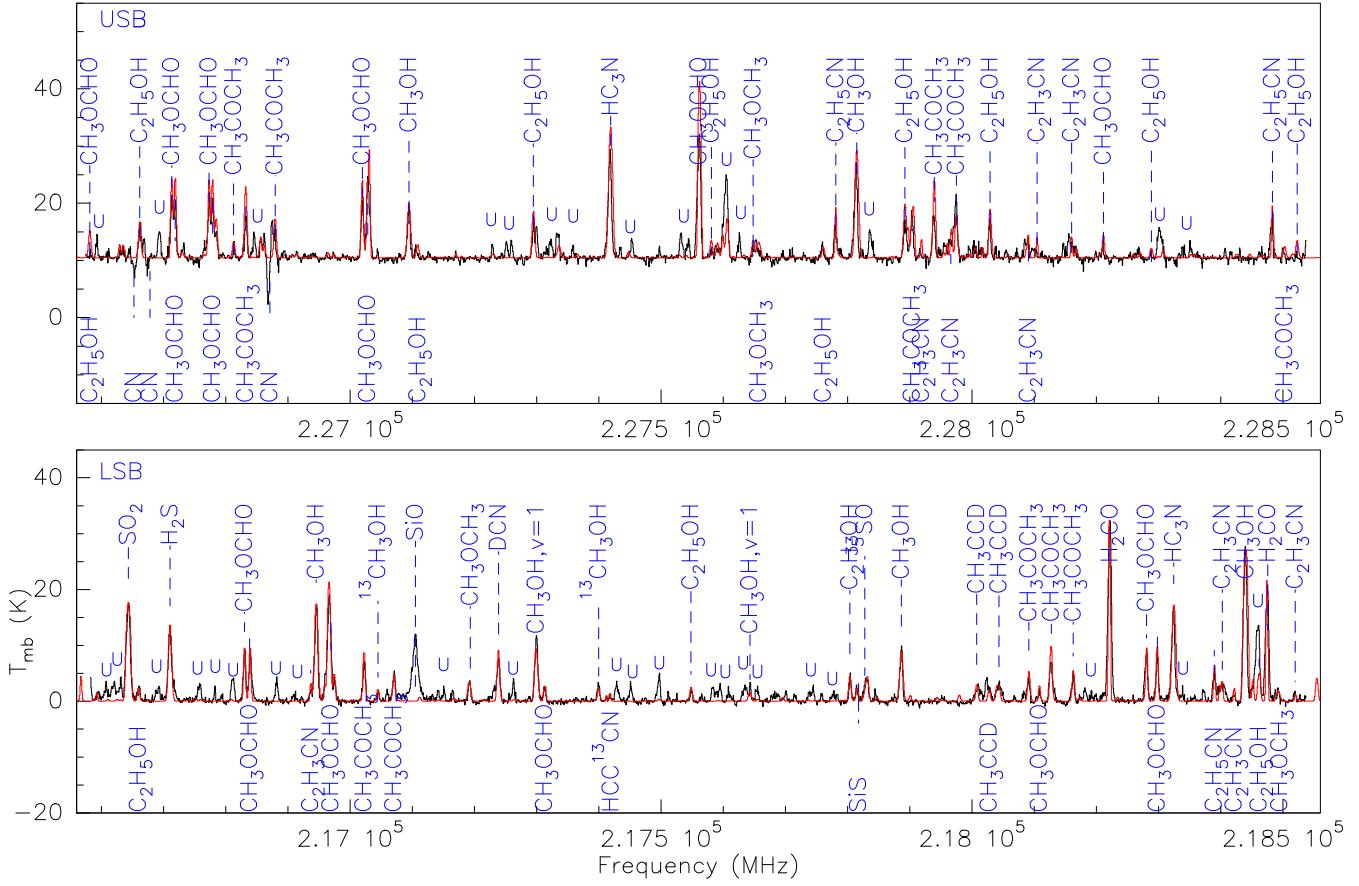}
\caption{The LSB and USB spectra extracted at the continuum peak position. The
  black curve is the observed spectra, and red curve indicates simulated
  spectra using the XCLASS program. U marks the unidentified transitions.}

\end{figure*}

\begin{figure*}
\hspace{-3cm}
\vspace{-1cm}
\includegraphics[angle=-90,width=1\textwidth]{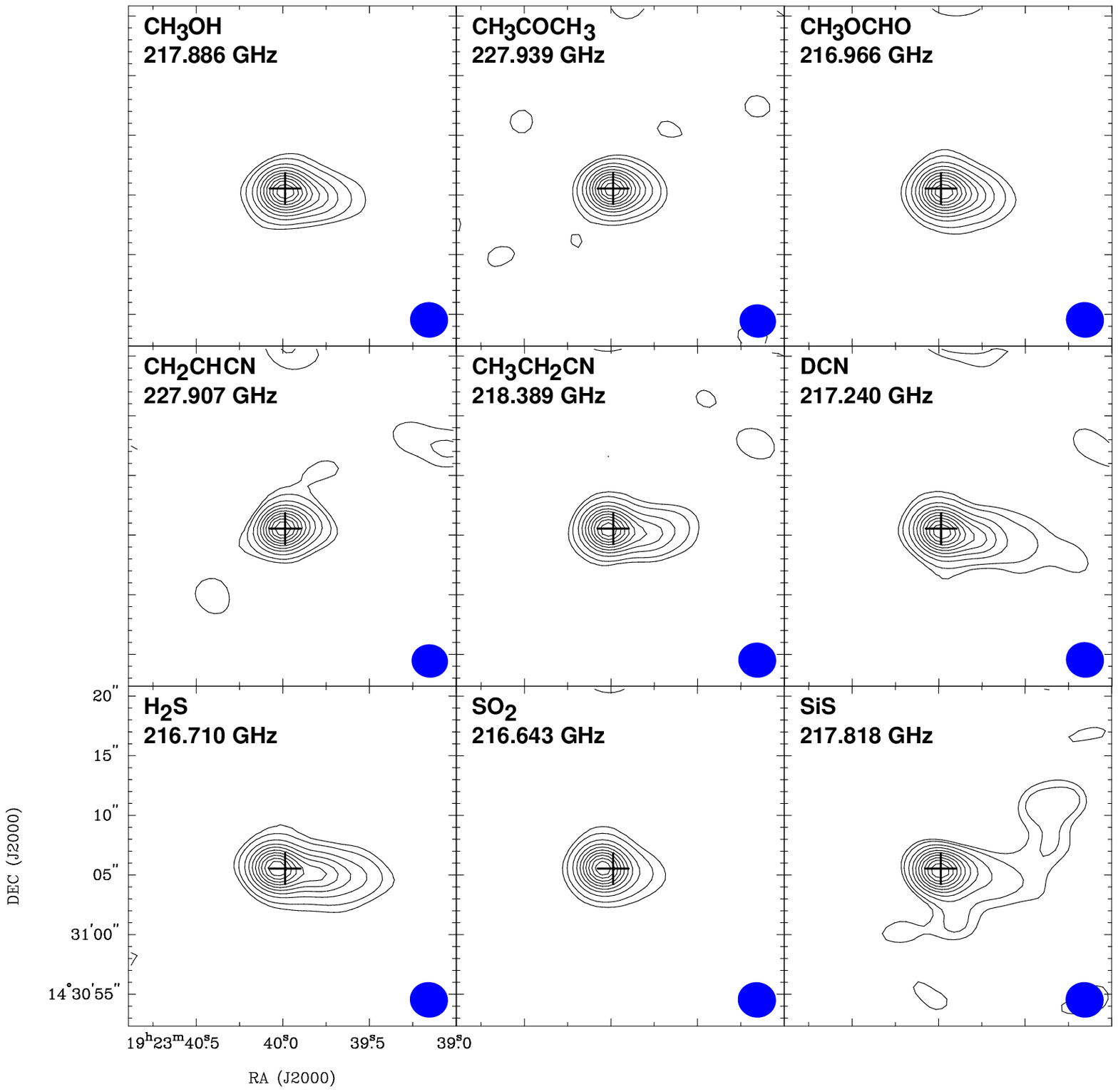}

\caption{The sample images of specific molecules. In each pannel, the
  synthesized beam is shown in the bottom right corner, the cross symbol marks
   the continuum peak position.  The contour levels are 5\%, 10\%, 20\%, 30\%,
   40\%, 50\%, 60\%, 70\%,80\%, 90\%, 100\% of the peak value. The peak values
   of CH$_3$OH, CH$_3$COCH$_3$, C$_2$H$_5$OH, CH$_2$CHCN, CH$_3$CH$_2$CN, DCN,
   H$_2$S, SO$_2$ and SiS are 24.7, 14.9, 78.6, 15.7, 14.1, 24.3, 45.4, 85.6
   and 95.2 Jy~beam$^{-1}$, respectively.}
\end{figure*}

\section{Data analysis}

The rotation temperatures and fractional abundances of complex molecules
can reflect immediate environments close to the stars or star-forming cores,
but also can provide key clues in understanding formation mechanism of the
specific molecules. In order to obtain reasonable parameters, we use the
XCLASS to estimate the column densities and rotation temperatures of the molecules.

\subsection{XCLASS analysis}

The XCLASS accesses the CDMS (M\"uller at al. 2001, M\"uller et
al. 2005; http://www.cdms.de) and JPL to get necessary molecular
essential parameters (Pickett et al. 1998; http://spec.jpl.nasa.gov). Under
the assumption of local thermodynamic equilibrium (LTE), the XCLASS
takes source size, beam filling factor, line profile, line
blending, background temperature, excitation and opacity into
account when solving radiative transfer equation. The detailed fitting functions and modeling
procedures are described in papers by Zernickel et al. (2012) and M\"oller,
Endres \& Schilke (2015). The source sizes
of these species are measured by use of two dimension Gaussian fitting to the
gas emission of specific molecule. In the LTE calculation, we fix the line
width and peak velocity unchanged, and set different rotation temperatures
and column densities to simulate the observed spectra. The
reasonable parameters were determined when the simulated spectra
have same line profiles as the observed ones (see red curve of Figure 2).

The column densities and rotation temperatures of 17 species and
their isotopologues are obtained from our LTE calculation, and the
results are summarized in Table 2.
Column (1) lists the molecular name, column (2) and (3) present
the column density and rotation temperature, respectively.

\setcounter{table}{1}
\begin{table*}
\caption{The Physical Parameters}
\begin{tabular}{lrrr} \hline

 Molecule&$N$&$T_{\rm rot}$& $f_{H_2}$\\
          &(cm$^{-2}$) &(K) &            \\
  \hline
CH$_3$OH  &$2.3\times10^{18}$&138&$3.6\times10^{-7}$\\
$^{3}$CH$_3$OH&$1.1\times10^{17}$&138&$1.7\times10^{-8}$\\
C$_2$H$_5$OH&$7.4\times10^{17}$&140&$1.2\times10^{-7}$\\
CH$_3$OCH$_3$&$1.8\times10^{17}$&140&$2.8\times10^{-8}$\\
CH$_3$COCH$_3$&$1.4\times10^{17}$&140&$2.2\times10^{-8}$\\
H$_2$CO&$4.3\times10^{16}$&110&$6.7\times10^{-9}$\\
CH$_3$OCHO&$3.0\times10^{17}$&138&$4.7\times10^{-8}$\\
&&&\\
CH$_2$CHCN&$9.2\times10^{15}$&140&$1.4\times10^{-9}$\\
CH$_3$CH$_2$CN&$2.6\times10^{16}$&140&$4.1\times10^{-9}$\\
HC$_3$N&$2.8\times10^{15}$&105&$4.4\times10^{-10}$\\
HCC$^{13}$CN&1.2$\times10^{14}$&105&1.9$\times10^{-11}$\\
&&&\\
SO$_2$&$9.4\times10^{17}$&180&$1.5\times10^{-7}$\\
$^{33}$SO&$3.4\times10^{15}$&170&$5.3\times10^{-10}$\\
H$_2$S&$4.4\times10^{16}$&143&$6.9\times10^{-9}$\\
SiS&$1.9\times10^{15}$&130&$3.0\times10^{-10}$\\

&&&\\
CH$_3$CCD&$6.6\times10^{16}$&130&$1.0\times10^{-8}$\\
DCN&$6.8\times10^{14}$&138&$1.1\times10^{-11}$\\
\hline
\end{tabular}
\end{table*}

From Table 2, the rotation temperatures of oxygen-bearing
molecules range from 110  to 140 K, while nitrogen-bearing
molecules have  rotation temperatures  ranging from 105  to 140 K,
which is consistent with that derived from the
transitions of CH$_3$CN (Zhang, Ho \& Ohashi 1998). The similar gas
temperatures and  distributions between oxygen- and
nitrogen-bearing molecules (see \S 3.3) suggest that oxygen-bearing
and nitrogen-bearing molecules are well mixed and originate from
the same physical environments. The scenario is inconsistent with the
observations toward the massive star-forming regions of Orion-KL
and G19.61-0.23, in which nitrogen-bearing molecules have higher
temperatures than oxygen-bearing molecules and oxygen- and nitrogen-bearing
molecules peak at different positions (Blake et al. 1987; Qin et al. 2010). The
sulfur-bearing molecules have gas temperatures ranging from 130
to 180 K. The sulfur-bearing molecules have been proved to be a
probe of outflows or shocks. The outflow motions are identified in W51 North
region (Zapata et al. 2008). The higher gas temperature of the sulfur-bearing
molecules in W51 North may be caused by outflow heating.

There are obvious difference in the column densities between the
oxygen-bearing and nitrogen-bearing molecules (see Table 2).  The column
densities of oxygen-bearing molecules range from $4.3\times10^{16}$ to
$2.3\times 10^{18}$ cm$^{-2}$. While the column densities of the
nitrogen-bearing molecules  range from $2.8\times 10^{15}$ to $2.6\times
10^{16}$ cm$^{-2}$.

\subsection{Abundance and isotopic ratio}

The fractional abundance of the observed molecules relative to H$_{2}$ are
 estimated by use of the expression, $f_{\rm H_{2}}=N_{T}/N_{\rm H_{2}}$, and
 summarized in column (4) of Table 2, where $N_T$ is column density of specific
 molecule and $N_{\rm H_{2}}$ is the column density of H$_2$. The
 oxygen-bearing molecules have relatively
higher abundances at a range of $6.7\times10^{-9}$ to $3.6\times10^{-7}$, while
the nitrogen-bearing molecules have abundances ranging from
$4.4\times10^{-10}$ to $4.1\times10^{-9}$. Previous observations of
massive star-forming regions also showed that the oxygen-bearing molecules have
higher abundances but lower temperatures than the nitrogen-bearing molecules
(e.~g., Blake et al. 1987; Wyrowski et al. 1999; Qin et al. 2010). The chemical
model proposed by Rodgers \& Charnley (2001) suggested that hot-core
composition (O-rich or N-rich) depends on core temperature and the
ammonia/methanol ratio of the evaporated ices, and the timescale of the core
formation, in which  O-rich cores form first, then become N-rich as
temperature rises and  evolution. In our case, the oxygen- and
nitrogen-bearing molecules peak at same position and have similar temperature,
which may be caused by the fact that W51 North is at an early
 evolutionary  stage and the hot core is not very developed, as suggested by
 dynamical analysis (Zhang, Ho \& Ohashi 1998; Zapata et al. 2008, 2009; Zapata, Tang \& Leurini 2010).

The CH$_3$OH, HC$_3$N and their isotopologues $^{13}$CH$_3$OH,
HCC$^{13}$CN are detected in W51 North. $^{12}$C/$^{13}$C
ratios of 21.0 and 23.3 are estimated from
CH$_3$OH/$^{13}$CH$_3$OH and HC$_3$N/HCC$^{13}$CN, respectively.
Based on the relationship of isotope ratios and distance from the
Galactic center by Willson \& Rood (1994), the $^{12}$C/$^{13}$C
ratio of 22 is obtained if one takes the distance of 1.5 kpc
to the Galactic center. The consistent
$^{12}$C/$^{13}$C ratio from the two methods suggest that the
parameters derived from the XCLASS modeling are reasonable.

\section{Individual molecules}
\subsection{Methanol (CH$_3$OH)}

Methanol (CH$_3$OH) is a key molecule in the chemical networks linked with the
formation of more complex oxygen-bearing organic molecules (Bottinelli et al.
2007; Whittet et al. 2011). Five rotational transitions of methanol
have been identified in our observations spanning an upper level energies of
$45-557$ K. In addition, we have detected two transitions of CH$_3$OH from
vibrational state at frequencies 217.29920 and 217.64286 GHz with upper level
energies of 374 and 746 K, respectively. The CH$_3$OH has highest fractional
abundance of $3.6\times 10^{-7}$ than other molecules. Higher CH$_3$OH
abundance is also observed in other massive star-forming regions (e.~g., Qin
et al. 2010, 2015; Neill et al. 2014).  The higher  CH$_3$OH abundance of
$\sim 10^{-7}$ can not be explained by gas-phase chemical model (Lee,
et al. 1996). The gas-phase chemical model suggested that
CH$_3$OH can be produced in gas phase via radiative association of CH$_3^+$
and H$_2$O at temperature $\leq$ 100 K, giving CH$_3$OH  abundance of $\sim$
$10^{-11}$ (Lee et al. 1996). In contrast, the grain chemical
model suggested that  CH$_3$OH is formed on grain surface via hydrogenation
of CO, followed with evaporating from grain surface into gas phase due to
temperature increasing, which can produce higher CH$_3$OH abundances.
The higher abundance of $\sim10^{-7} $ in our observations favor that this molecule originates from grain
surface chemistry (Charnley et al. 2004; Garrod \& Herbst 2006).

\subsection{Acetone (CH$_3$COCH$_3$)}
 Acetone (CH$_3$COCH$_3$) had been detected for the first time by Combes et al. (1987) and later
 confirmed by Snyder et al. (2002). So far acetone (CH$_3$COCH$_3$) was only
 detected in the hot molecular core Sgr B2(N) (Snyder et al. 2002), the Orion
 KL hot core (Friedel et al. 2005; Peng et al. 2013), and G24.78+0.08  (Codella
 et al. 2013). We have identified 13 transitions of CH$_3$COCH$_3$ in W51
 North region, spanning a wide energy range of $66-418$ K. The column density of
 $1.4 \times 10^{17}$cm$^{-2}$ and gas temperature of 140 K are estimated by use
 of the XCLASS, and the higher abundance of $2.0\times 10^{-8}$ is obtained.

 Combes et al. (1987) proposed that CH$_3$COCH$_3$ is formed in gas-phase
  via the radiative association reaction. However Herbst, Giles \& Smith (1990) argued
  that this radiative association reaction is very inefficient to produce
 observed CH$_3$COCH$_3$ abundance, and this process can only explain the lower
 abundance of $\sim 10^{-11}$. Recently, the observations of CH$_3$COCH$_3$
 toward the Orion-KL have shown that the distribution of CH$_3$COCH$_3$ is
 similar to N-bearing molecules concentrated at the hot core rather than
 O-bearing molecules peaked at the Compact ridge, therefore the
 formation of CH$_3$COCH$_3$ may invoke N-bearing molecules, and grain surface
 chemistry or high temperature gas-phase chemistry may play important roles
 (Friedel et al. 2005, 2008; Peng et al. 2013). The well mixed nitrogen- and
 oxygen-bearing molecules based on gas distributions and
 temperatures in W51 North indicate that the hot core is not very developed in
 W51 North and grain surface chemistry and evaporating from grain surface into
 gas phase may be
 responsible for the high CH$_3$COCH$_3$ abundance of $2.0\times 10^{-8}$,
 since the gas temperature is not so high in W51 North.

 Compared to  the Sgr B2(N) (Snyder et al. 2002), the Orion-KL (Peng et
   al. 2013), and  G24.78+0.08 (Codella et al. 2013) molecule line cores, W51 North region has
   higher CH$_3$COCH$_3$ abundance. The CH$_3$COCH$_3$ can be destructed in
 gas phase by efficient
 radiative association of H$_3^+$ and CH$_3$COCH$_3$ or the collisions of
CH$_3$COCH$_3$ and C$^+$ (Combes et al. 1987; Herbst, Giles \& Smith 1990). A
 possible  interpretation of higher CH$_3$COCH$_3$ abundance is that unlike the
 Sgr B2(N), the Orion-KL, and G24.78+0.08, W51 North is at very
 early stage of massive star formation, and the CH$_3$COCH$_3$ is destructed
 less  and converted into the other molecules in a short time.

\subsection{Ethanol (C$_2$H$_5$OH)}

We have identified 13 transitions of C$_2$H$_5$OH in W51 North region.
These transitions have a wide spread of the upper level energies of $23-305$ K.
The C$_2$H$_5$OH has higher abundance of $1.2 \times 10^{-7}$ in our
observations. Compared to other sources, the C$_2$H$_5$OH abundance of $1.2\times 10^{-7}$ and
$N_{\rm C_{2}H_{5}OH}$/$N_{\rm CH_{3}OH}$ $\approx$ 0.32 are much higher than those
in other massive star-forming regions (Ohishi et al. 1995; Qin et al. 2010;
Bisschop et al. 2007). Hence W51 North is a very ethanol-rich source.

  The gas-phase chemical reaction via ion-molecule reaction predicted
 C$_2$H$_5$OH abundance of $\leq 10^{-11}$ (Leung, Herbst \& Huebenr
 1984). Higher  abundance of $1.2 \times 10^{-7}$ in our observations can't be
 explained by pure gas-phase chemical model. The grain chemistry suggested
 that C$_2$H$_5$OH can be formed on the grain surface via hydrogenation of
 CH$_3$CO  (Charnley et al. 2004; Ohishi et al. 1995), and the
 model predicted high C$_2$H$_5$OH abundance of $10^{-8} \sim 10^{-7}$
 within $\sim$ $10^{4}$ yr after evaporated from the grain surface into gas
 phase. Our result appears to support that the C$_2$H$_5$OH is synthesized
 on grain surface (Ohishi et al. 1995; Charnley et al. 1995).

\subsection{Dimethyl ether (CH$_3$OCH$_3$) and Methyl formate (CH$_3$OCHO)}

Four transitions of CH$_3$OCH$_3$ are detected in W51 North. We
estimated its rotation temperature of 140 K and column density of
$1.8\times10^{17}$ cm$^{-2}$ from the XCLASS calculation. The abundance ratio of
$N_{\rm CH_{3}OCH_{3}}$/$N_{\rm C_{2}H_{5}OH}$ is approximately 0.24. The
$N_{\rm CH_{3}OCH_{3}}$/$N_{\rm C_{2}H_{5}OH}$ ratios have a great deal of
changes from source to source (Fontani et al. 2007; Qin et al.
2010; Millar et al. 1988; Schilke et al. 1997; Bisschop et al.
2007). Although CH$_3$OCH$_3$ is isomer of C$_2$H$_5$OH, the
abundances of C$_2$H$_5$OH and CH$_3$OCH$_3$ are not well correlation,
suggesting that they have different chemical route, and do not have same
'parent' molecules.

 As one of large oxygen-bearing molecules, CH$_3$OCHO has been
reported in many hot molecular cores and corinos (Blake et al. 1987;
Hatchell et al. 1998; Nummelin et al. 2000; Cazaux et al. 2003; Bottinelli
et al. 2004). Sixteen transitions of CH$_3$OCHO are identified in W51 North,
spanning a upper energy level range of $100-358$ K. High fractional
abundance of $4.7\times 10^{-8}$ is
obtained. Gas temperature, column density and spatial distribution
of CH$_3$OCHO are similar to those of CH$_3$OCH$_3$ (see Table 2
and Figure 3). Similar case is also seen in Orion-KL, in which the
distributions of CH$_3$OCHO and CH$_3$OCH$_3$ present a striking
similarity, and the two species have comparable gas temperature
and column density (Brouillet et al. 2013), suggesting that
CH$_3$OCHO and CH$_3$OCH$_3$ may be chemically related and have
similar formation mechanism (Brouillet et al. 2013).

 Charnley, Tielens \& Millar (1992) proposed that CH$_3$OCH$_3$ and CH$_3$OCHO
can be formed in gas phase chemistry via molecule-ion reaction of CH$_3$OH$_2^+$
 with CH$_3$OH and H$_2$CO, respectively. Their model predicted high abundance
 of $\sim 10^{-8}$ for CH$_3$OCH$_3$ and CH$_3$OCHO at age of $\sim10^{4}$ yr.
Our results appear to favor the gas-phase formation routes of CH$_3$OCH$_3$ and
 CH$_3$OCHO by Charnley, Tielens \& Millar (1992).

\subsection{Vinyl cyanide (CH$_2$CHCN) and Ethyl cyanide (CH$_3$CH$_2$CN)}

We have identified 9 CH$_2$CHCN transitions with upper level energies E$_u$
 of $134-341$ K. The column density of $9.2\times 10^{15} $cm$^{-2}$ and the
 rotation temperature of 140 K are obtained. Three CH$_3$CH$_2$CN
 transitions with upper level energies E$_u$ of $139-151$ K are identified. The
 column density of $2.6\times 10^{16} $ cm$^{-2}$ and rotation temperature of
 140 K are estimated by use of the XCLASS.

CH$_2$CHCN  and CH$_3$CH$_2$CN have same temperature of 140 K, while
CH$_3$CH$_2$CN has higher column density than CH$_2$CHCN. Previous
observations suggested that the abundances of two molecules  are strongly
correlated (Fontani et al. 2007). $N_{\rm CH_{2}CHCN}$/$N_{\rm CH_{3}CH_{2}CN}$ in W51
North is approximate 0.35 which agrees well with abundance  correlation of
CH$_3$CH$_2$CN and CH$_2$CHCN in other star-forming regions (Fontani et
al. 2007)  and CH$_2$CHCN may
be formed through gas phase reactions involving CH$_3$CH$_2$CN (Caselli,
Hasegawa \& Herbst 1993). While CH$_3$CH$_2$CN can be formed
 by successive hydrogenation of HC$_3$N on dust grains and released into the
 gas phase as temperature rises (Blake et al. 1987; Caselli, Hasegawa \&
 Herbst 1993). The ratio of $N_{\rm CH_3CH_2CN}$/$N_{\rm HC_{3}N}$
 $\approx$ 10 in our observations is consistent with the model prediction of
 grain surface chemistry (Blake et al. 1987).

\subsection{Deuterated molecules }

We have detected one transition of the deuterated species of hydrogen cyanide
(DCN), in W51 North. The column
density of $6.8\times10^{14}$ cm$^{-2}$ is estimated, which is
much larger than those in other sources associated with UC H{\sc
ii} regions (Hatchell, Millar \& Rodgers 1998). The deuterated species are
thought to be synthesized at the early evolutionary stage of star formation
(Miettinen, Hennemann \& Linz 2011).  Previous studies suggested that
DCN can be formed via D-H substitution of the HCN or the reaction of CHD with
N on the grain mantles and then be released
into the gas-phase, while DCN can be destroyed with temperature increasing
(Schilke et al. 1992; Hatchell, Millar \& Rodgers 1998).  Hence the larger
abundance observed in W51 North is not a surprise and suggests that it may be
synthesized at the early evolutionary stages of star formation.

As isotopomer of methyl acetylene CH$_2$DCCH has been reported in
the dark cloud TMC-1 CP by Gerin et al. (1992), and CH$_3$CCD had
been successfully detected in same region by Markwick et al.
(2005). So far the CH$_3$CCD has been less reported in other
massive star formation regions. Therefore CH$_3$CCD is an important
molecule for investigating the difference of
physical and chemical environments between the dark clouds and hot
molecular cores. We have successfully identified 3 transitions
of CH$_3$CCD with E$_u$ ranging from 78 to 143 K. Higher CH$_3$CCD
abundance of $1\times10^{-8}$ and temperature of
130 K  compared to dark clouds (Markwick et al. 2005) are
estimated, which may indicate that CH$_3$CCD is synthesized on
grain surface and release to gas phase as temperature increases.

\subsection{Silicon monosulfide (SiS)}
 One transition of SiS (J=12--11) at 217.81766 GHz is
 identified, and the estimated column density and abundance are $1.9\times
 10^{15}$ cm$^{-2}$ and $3.0\times 10^{-10}$, respectively.
 So far this species is only detected in carbon star IRC+10216, Sgr B2(N),
 Sgr B2(M) and Sgr B2(OH) (Bieging \& Nguyen 1989; Turner 1991; Belloche et
 al. 2013).  The column density of  $1.9\times 10^{15}$ cm$^{-2}$ is
 higher than $\sim 10^{13}$cm$^{-2}$ in Sgr B2. While the factional
 abundance of $3.0 \times 10^{-10}$ in W51 North is lower than $\sim
 10^{-6}$ in the carbon star, IRC+10216. The abundance difference of four
 order of magnitude  may be caused by the different physical and chemical
environments in massive star formation regions and the asymptotic giant branch star.

\subsection{Sulfur-bearing molecules}
Sulfur-bearing molecules of H$_2$S and SO$_2$ are observed in W51
North, with higher gas temperature than other molecules. The
fractional abundances of SO$_2$ and H$_2$S are $1.5\times10^{-7}$
and $6.9\times10^{-9}$, respectively. The ratio of
$N_{\rm H_2S}$/$N_{\rm SO_2}$  is appropriately 0.05, which is same order
of magnitude as those in massive protostar cores (Herpin et al.
2009), which is consistent with chemical model that H$_2$S can be formed
  on the grain surface and H$_2$S is converted to SO first, then
the SO is converted to SO$_2$ as temperature rises (Charnley 1997;
Wakelam et al. 2004; Woods et al. 2015).

\section{SUMMARY}

 We present the Submillimeter Array (SMA) observations of molecular lines
 in two 2 GHz-wide bands centered at 217.5 and 227.5 GHz, toward massive
 star forming region W51 North. We identified 84 transitions from 17 species,
 including oxygen-, nitrogen- and sulfur-bearing molecules.
 Our main conclusions are as follows: \\

\noindent  1. The gas distributions of both oxygen-bearing and
nitrogen-bearing molecules show a compact core concentrated on the strongest continuum
source, which indicates that nitrogen- and oxygen-bearing molecules are
well-mixed in space in W51 North region. \\

\noindent  2. Under the assumption of local thermodynamic equilibrium, the
molecular rotation temperatures and column densities are estimated by use of
the XCLASS program. The oxygen-bearing molecules have higher
fractional abundance than the nitrogen-bearing molecules. The rotation
temperatures range from 100 to 200 K, suggesting that the molecular
emissions originate from warm environments.   \\

\noindent  3. Thirteen transitions of CH$_3$COCH$_3$ are identified. Higher
fractional abundance of CH$_3$COCH$_3$ are obtained in W51 North when compared
to the other massive star-forming regions, {\it e.g.} Sgr B2(N), Orion KL,
G24.78+0.08.  These  results seem to indicate that CH$_3$COCH$_3$ is
synthesized on grain surface at the early evolutionary stage of massive star formation.  \\

\noindent  4. Higher fractional abundance of CH$_3$OH and C$_2$H$_5$OH are
obtained, which cannot explained by gas phase reactions and  the two molecules
may be synthesized on grain surface. $N_{C_{2}H_{5}OH}$/$N_{CH_{3}OH}$
$\approx$ 0.32 are much higher than those in other massive
star-forming regions, and W51 North is a very ethanol-rich source. \\

\noindent 5. Similar gas distribution, rotation temperature and abundance
between CH$_3$OCHO and CH$_3$OCH$_3$ suggest that they may be chemically related
and  have similar formation mechanism.  The two molecules may originate
  from gas phase chemistry.     \\

\noindent  6. $N_{\rm CH_{2}CHCN}$/$N_{\rm CH_{3}CH_{2}CN}$  in W51 North
 agrees well with abundance correlation
of CH$_{3}$CH$_{2}$CN and CH$_{2}$CHCN in other star-forming
regions. CH$_{3}$CH$_{2}$CN   can be formed
by successive hydrogenation of HC$_{3}$N on dust grains while  CH$_{2}$CHCN may be formed
through gas phase reactions involving CH$_{3}$CH$_{2}$CN.  \\

\noindent 7. CH$_3$CCD and SiS have been less reported in other star-forming
regions. Higher fractional abundances of CH$_3$CCD and SiS are
estimated in W51 North region.


\bigskip

\section*{Acknowledgments}
 We thank the anonymous referee, and editor Morgan Hollis for their
 constructive comments on the paper. This work has been supported by the
 National Natural Science Foundation of China under grant Nos. 11373026,
 11373009, 11433004, 11433008, U1331116, and the National Basic Research
Program of China (973 Program) under grant No. 2012CB821800, by Top Talents
Program of Yunnan Province and Midwest universities comprehensive strength
promotion project (XT412001, Yunnan university).


 \end{document}